\documentclass[11pt,a4paper]{article}

\usepackage[utf8]{inputenc}
\usepackage[T1]{fontenc}
\usepackage{amsmath,amssymb,amsthm}
\usepackage{geometry}
\usepackage{comment}
\usepackage{graphicx}

\usepackage[colorlinks=true,linkcolor=blue,citecolor=blue,urlcolor=blue]{hyperref}
\usepackage{enumitem}
\hypersetup{
  colorlinks=true,
  linkcolor=blue,
  urlcolor=blue,
  citecolor=blue
}

\usepackage{url}

\title{Time-resolved digital quantum simulation of cosmological particle creation in a de Sitter–radiation transition.}

\author{
Hamzeh Alavirad \textsuperscript{} \\[0.5em]
\small \textsuperscript{}QUSMOS \\ Karlsruhe, Germany \\
\small \texttt{h.alavirad@qusmos.com}
}

\date{\today}

\begin{document}

\maketitle

\begin{abstract}
We present a time-resolved digital quantum simulation of cosmological particle
creation in a de~Sitter--radiation FLRW transition. Instead of compiling only
the final Bogoliubov transformation into a one-shot circuit, we discretize the
conformal-time evolution and implement the dynamics as a Trotterized sequence
of short-time circuit blocks. This formulation gives access not only to the
late-time particle number, but also to the build-up of fixed-basis pair
occupation during the non-adiabatic transition. Using a four-qubit
single-excitation encoding for a momentum pair \((+\mathbf{k},-\mathbf{k})\),
we compare matrix-Trotter evolution, noiseless statevector simulation,
finite-shot Qiskit Aer simulation, and a shallow \(N=1\) IBM hardware
implementation. The simulator results are consistent with the analytic sudden-transition
benchmark \(n_k=1/[4(k\eta_e)^4]\) in the controlled single-excitation
regime. The IBM experiment demonstrates
execution of the shallow circuit block, but exhibits a residual hardware error
of order \(10^{-2}\), indicating that quantitative hardware reconstruction of
the particle spectrum remains beyond current NISQ performance.
\end{abstract}


\section{Introduction}

Feynman's proposal that quantum systems could be used to simulate other quantum systems~\cite{Feynman1982} laid the conceptual foundation for the modern field of quantum simulation, which now includes both analog platforms and gate-based digital approaches~\cite{Georgescu2014,Altman2021}. 
In the current noisy intermediate-scale quantum (NISQ) regime, quantum
simulation remains one of the most relevant near-term applications of quantum
processors, especially for real-time unitary dynamics, where Hamiltonian time
evolution is naturally represented by quantum circuits~\cite{Lloyd1996,Preskill2018}.

A broad range of physics problems have been proposed as targets for quantum
processors. In quantum chemistry, early algorithmic demonstrations showed how
molecular energies can be estimated with quantum resources in ways that avoid
the steep scaling of exact classical methods~\cite{AspuruGuzik2005}. In
high-energy and field-theory settings, quantum algorithms for relativistic
quantum field theories have been developed in principle~\cite{JordanLeePreskill2012},
and lattice gauge theories have become an important target for quantum
simulation~\cite{Banuls2020}. Experimentally, real-time gauge-theory dynamics
have already been implemented on small quantum devices, for example in digital
simulations of $(1+1)$-dimensional QED and Schwinger pair production
\cite{Martinez2016}. These examples are especially relevant for real-time
unitary dynamics, where quantum devices provide a natural computational
framework.

A particularly natural target for quantum simulation is quantum field
theory in time-dependent backgrounds. In quantum field theory in curved
spacetime, the notion of a particle depends on the choice of positive-frequency mode basis:
different mode decompositions define different vacua, and a time-dependent
geometry can mix positive- and negative-frequency components through a
Bogoliubov transformation~\cite{BirrellDavies1982,ParkerToms2009}. In a
spatially flat FLRW universe, the rescaled scalar-field modes obey a
harmonic-oscillator equation with a time-dependent effective frequency. When
the evolution is non-adiabatic, early-time ``in'' modes and late-time ``out''
modes are related by nontrivial Bogoliubov coefficients. In homogeneous and
isotropic backgrounds, momentum conservation implies that quanta are produced
in correlated pairs with opposite comoving momenta
$(+\mathbf{k},-\mathbf{k})$. This two-mode structure is the field-theoretic
origin of cosmological squeezing~\cite{Mukhanov2005} and makes cosmological
particle creation well suited to a digital quantum-simulation description in a
truncated Fock space.

Gate-based studies have already begun to address curved-spacetime and
gravity-inspired settings. Maceda and Sab\'in performed a digital quantum
simulation of cosmological particle creation on IBM quantum processors,
estimating the produced particle number and studying the role of error
mitigation~\cite{MacedaSabin2025}. Related IBM-based demonstrations have
implemented quadratic bosonic Hamiltonians using boson--qubit mappings, gate
decompositions, and error-mitigation techniques in gravity-inspired
optomechanical and entanglement-generation settings
\cite{CarmonaRufoEtAl2024Opto,Sabin2023GravEnt}. A related fermionic study has
treated fields in expanding spacetimes using Jordan--Wigner mappings and
Trotterized evolution~\cite{GongYang2025}. Outside the gate-based setting,
complementary proposals connect curved-spacetime quantum field theory to
controllable many-body systems and analog quantum devices
\cite{KinoshitaMurata2025,TerronesSabin2021}.

Compared with the one-shot IBM implementation~\cite{MacedaSabin2025}, where
the circuit is compiled from the final analytic Bogoliubov coefficients, the
present paper constructs the evolution from the time-dependent Hamiltonian coefficients. The time-ordered evolution
is approximated by a sequence of short conformal-time steps, each corresponding
to a circuit block with rotation angles determined by the instantaneous
Hamiltonian coefficients. This allows the non-adiabatic build-up of the
fixed-basis pair occupation to be followed through the transition and makes
the method applicable to backgrounds for which the final Bogoliubov
coefficients are not available in closed form. The cost is a larger circuit
depth.

The resulting circuit depth grows with the number of time slices. We therefore
validate the full large-\(N\) construction on quantum-circuit simulators, while
the IBM hardware experiment is restricted to a shallow \(N=1\) representative
block.

The paper is organized as follows. Section~\ref{sec:ds_radiation_model}
introduces the de~Sitter--radiation toy model, derives the analytic
Bogoliubov benchmark, and discusses the validity of the single-excitation
regime. Section~\ref{sec:digital_simulation} formulates the time-resolved
Hamiltonian evolution, the conformal-time discretization, and the four-qubit
encoding. Section~\ref{sec:results} presents the simulator validation,
finite-shot sampling results, and the shallow IBM hardware demonstration.
Section~\ref{sec:conclusion} summarizes the conclusions and limitations.

\section{de Sitter--Radiation Model}
\label{sec:ds_radiation_model}

We consider a spatially flat FLRW spacetime whose evolution is idealized
as a transition from an early de Sitter phase to a later radiation-dominated
phase. In conformal time, the metric is
\begin{equation}
  ds^2 = a^2(\eta)\left(-d\eta^2+d\mathbf{x}^2\right),
  \label{eq:flrw_metric}
\end{equation}
where \(a(\eta)\) is the scale factor.

The transition is assumed to occur around a conformal time
\(\eta_e<0\). We approximate the transition by a sudden matching, in which the scale factor
and its first derivative are continuous at \(\eta_e\), while \(a''/a\)
changes discontinuously. In this sudden-transition model, the scale factor is chosen as
\begin{equation}
  a(\eta)=
  \begin{cases}
    -\dfrac{1}{H\eta},
    & \eta\leq \eta_e, \\[1.2em]
    a_e\left(2-\dfrac{\eta}{\eta_e}\right),
    & \eta\geq \eta_e,
  \end{cases}   
  \label{eq:piecewise_scale_factor}
\end{equation}
where \(a_e\equiv a(\eta_e)=-{1}/{(H\eta_e)}.\) The first branch describes de Sitter expansion with approximately constant
Hubble parameter \(H\). The second branch is linear in conformal time, as
appropriate for radiation domination. This choice makes both \(a(\eta)\)
and \(a'(\eta)\) continuous at the transition, however, \(a''(\eta)\) is discontinuous at \(\eta=\eta_e\). This discontinuity should not be interpreted as a physical discontinuity in the
reheating dynamics. It is used as the sudden-transition approximation to a rapid
but finite change from inflation to radiation domination. This approximation
allows the Bogoliubov coefficients to be computed analytically and therefore
provides a controlled benchmark for the quantum-simulation results.

We now add to this background a real, massless, minimally coupled scalar
field \(\phi\). Defining the rescaled canonical field \(v(\eta,\mathbf{x}) = a(\eta)\phi(\eta,\mathbf{x})\)
and Fourier decomposing into comoving modes, each mode satisfies
\begin{equation}
  v_k''(\eta)+\omega_k^2(\eta)v_k(\eta)=0,
  \qquad
  \omega_k^2(\eta)
  =
  k^2-\frac{a''(\eta)}{a(\eta)},
  \label{eq:mode_equation}
\end{equation}
where \(k=|\mathbf{k}|\). From Eq.~\eqref{eq:piecewise_scale_factor}we have
\begin{equation}
  \omega_k^2(\eta)=
  \begin{cases}
    k^2-\dfrac{2}{\eta^2},
    & \eta<\eta_e, \\[0.8em]
    k^2,
    & \eta>\eta_e .
  \end{cases}
  \label{eq:piecewise_frequency}
\end{equation}

Particle creation is caused by the non-adiabatic change of the effective
frequency \(\omega_k(\eta)\). In the sudden-transition approximatio model, this non-adiabaticity is
compressed into the matching surface \(\eta=\eta_e\), where \(a''/a\)
changes abruptly. In a smoother model, the same mechanism would arise from
a rapid but continuous variation of \(a''/a\).

We denote by \(\chi_k(\eta)\) the mode solution selected by the Bunch--Davies condition, i.e., the solution that approaches the positive-frequency Minkowski mode in the asymptotic de~Sitter past and is then continued through the transition~\cite{BunchDavies1978}. In the radiation era, where \(a''=0\), the same solution
is expanded in positive- and negative-frequency plane waves:
\begin{equation}
\chi_k(\eta)=
\begin{cases}
\dfrac{1}{\sqrt{2k}}
\left(1-\dfrac{i}{k\eta}\right)e^{-ik\eta},
& \eta<\eta_e, \\[1.2em]
\dfrac{1}{\sqrt{2k}}
\left(
\alpha_k e^{-ik\eta}
+
\beta_k e^{ik\eta}
\right),
& \eta>\eta_e .
\end{cases}
\label{eq:global_mode_solution}
\end{equation}

The coefficients \(\alpha_k\) and \(\beta_k\) are the Bogoliubov
coefficients relating the early-time de Sitter in-basis to the late-time
radiation out-basis.

Continuity of the global mode \(\chi_k\) and its first derivative at \(\eta=\eta_e\) gives 
\begin{equation}
  \alpha_k
  =
  1-\frac{i}{k\eta_e}
  -\frac{1}{2(k\eta_e)^2},
  \qquad
  \beta_k
  =
  \frac{e^{-2ik\eta_e}}{2(k\eta_e)^2}.
  \label{eq:bogoliubov_coefficients}
\end{equation}
These coefficients satisfy the normalization condition \(|\alpha_k|^2-|\beta_k|^2=1.\)

Because the background is spatially homogeneous, comoving momentum is
conserved. Particles are therefore produced in correlated pairs with
opposite comoving momenta, \(+\mathbf{k}\) and \(-\mathbf{k}\). At the
operator level, this pair structure is expressed by the Bogoliubov
transformation
\begin{equation}
  a_{\mathbf{k}}^{\rm out}
  =
  \alpha_k a_{\mathbf{k}}^{\rm in}
  +
  \beta_k^* a_{-\mathbf{k}}^{{\rm in}\dagger},
  \qquad
  a_{-\mathbf{k}}^{\rm out}
  =
  \alpha_k a_{-\mathbf{k}}^{\rm in}
  +
  \beta_k^* a_{\mathbf{k}}^{{\rm in}\dagger}.
  \label{eq:operator_bogoliubov}
\end{equation}

The expected number of out-particles in mode \(\mathbf{k}\), when the
field is initially in the de Sitter in-vacuum, is therefore
\begin{equation}
  n_k
  =
  \langle 0_{\rm in}|
  {a_{\mathbf{k}}^{\rm out}}^\dagger a_{\mathbf{k}}^{\rm out}
  |0_{\rm in}\rangle
  =
  |\beta_k|^2
  =
  \frac{1}{4(k\eta_e)^4}.
  \label{eq:sudden_particle_number}
\end{equation}
This quantity is an asymptotic late-time particle number defined with respect
to the radiation-era out-basis. It provides the analytic reference against
which the late-time output of the time-resolved quantum simulation is compared.

For the four-qubit digitization, each mode is truncated to the subspace with
at most one excitation. This single-excitation truncation is controlled only
when the expected occupation is small, \(n_k\ll 1\), or equivalently for
sufficiently large \(|k\eta_e|\). Parameter choices with \(n_k\sim 1\) or
larger require a higher Fock-space cutoff and lie outside the controlled
regime of the encoding used here. Increasing the cutoff would require more
qubits per mode, additional Pauli terms in the encoded Hamiltonian, and
deeper circuits, making such regimes more demanding for present noisy
hardware.

The sudden transition should be understood as an analytically tractable
limit of a rapid but smooth transition. In a more realistic model,
\(a''/a\) would vary continuously over a finite conformal-time interval
\(\Delta\), and the resulting Bogoliubov coefficients would generally have
to be obtained numerically. Such a smooth profile suppresses particle
creation for modes that evolve adiabatically across the transition,
approximately those with \(k\Delta \gg 1\). In this work we keep the
sudden-transition model because it provides closed-form Bogoliubov
coefficients and therefore a clean benchmark for the quantum simulation.

\section{Digital Quantum Simulation}
\label{sec:digital_simulation}
In this section we convert the mode dynamics of
Sec.~\ref{sec:ds_radiation_model} into a digitally implementable form. Each
momentum pair \((+\mathbf{k},-\mathbf{k})\) is written as a quadratic
two-mode system in a fixed Fock basis, so that the background evolution
appears through time-dependent Hamiltonian coefficients. The resulting
time-ordered evolution is then discretized and mapped to a four-qubit
encoding.

\subsection*{Fixed ladder operators and two-mode generators}

It is convenient to describe the dynamics in terms of \emph{fixed} creation and
annihilation operators for the mode pair $(+k,-k)$, defined with respect to a
time-independent reference frequency. Since in the radiation era the effective
frequency tends to $k$, the reference frequency is chosen as \(\omega_0 = k.\)

Time-independent ladder operators $a_k,a_k^\dagger$ and $a_{-k},a_{-k}^\dagger$
are then introduced in the usual way, and the dynamics is entirely transferred
into a time-dependent Hamiltonian $H_k(\eta)$ acting on this fixed Fock space.
For the pair \((+\mathbf{k},-\mathbf{k})\), we define the Hermitian operators
\begin{equation}
\begin{aligned}
  Z &\equiv a_k^\dagger a_k + a_{-k}^\dagger a_{-k},\\
  A &\equiv a_k a_{-k} + a_k^\dagger a_{-k}^\dagger .
\end{aligned}
\label{eq:ZA_def}
\end{equation}
The operator \(Z\) counts the total number of quanta in the pair, while
\(A\) generates the real two-mode squeezing interaction that creates and
annihilates correlated \((+\mathbf{k},-\mathbf{k})\) pairs. For a more
general complex squeezing coefficient, an additional Hermitian quadrature
\(B=i(a_k a_{-k}-a_k^\dagger a_{-k}^\dagger)\) would appear; in the real
FLRW background and phase convention used here its coefficient vanishes.

\subsection*{Time-dependent Hamiltonian}

The physical origin of the interaction terms in the Hamiltonian is seen most
clearly by starting from the field-theoretic description. For a real, massless,
minimally coupled scalar field, the action on the background
\eqref{eq:flrw_metric} is
\begin{equation}
  S = \frac{1}{2} \int d\eta\,d^3x\,a^2(\eta)
  \left[
    \phi'^2-(\nabla\phi)^2
  \right].
  \label{eq:scalar_action}
\end{equation}
After Fourier transforming and introducing the rescaled mode
\(v_k(\eta)=a(\eta)\phi_k(\eta)\), the Hamiltonian can be written as a sum over
independent parametric oscillators,
\begin{equation}
  H = \frac{1}{2}\sum_{\mathbf{k}}
      \Big(|\pi_k|^2 + \omega_k^2(\eta)\,|v_k|^2\Big),
  \label{eq:parametric_oscillator_hamiltonian}
\end{equation}
where \(\omega_k(\eta)\) is the conformal-time frequency defined in
Eq.~\eqref{eq:mode_equation}.

A time-dependent oscillator basis could in principle be introduced when the
instantaneous frequency is real and slowly varying. However, such a basis would
itself change with conformal time. For the digital simulation it is more useful
to work in a fixed Hilbert-space basis, so that the same qubit encoding is used
throughout the evolution.

We therefore work in the fixed Fock basis introduced above, with reference
frequency \(\omega_0=k\). In this basis, the states are time independent and
all background dependence appears in the coefficients of the Hamiltonian.

Expressing the parametric-oscillator Hamiltonian
\eqref{eq:parametric_oscillator_hamiltonian} in the fixed Fock basis introduces
off-diagonal terms. For the momentum pair \((+\mathbf{k},-\mathbf{k})\), we
write the fixed-basis canonical variables as
\[
  v_k =
  \frac{1}{\sqrt{2\omega_0}}
  \left(a_k+a_{-k}^\dagger\right),
  \qquad
  \pi_k =
  -i\sqrt{\frac{\omega_0}{2}}
  \left(a_k-a_{-k}^\dagger\right).
\]
For a real scalar field, the modes \(\mathbf{k}\) and \(-\mathbf{k}\) are
related by \(v_{-\mathbf{k}}=v_{\mathbf{k}}^*\) and
\(\pi_{-\mathbf{k}}=\pi_{\mathbf{k}}^*\). Grouping these contributions, the
Hamiltonian associated with one momentum pair can be written as
\[
  H_k=|\pi_k|^2+\omega_k^2(\eta)|v_k|^2 .
\]
Substituting the fixed-basis variables into this pair Hamiltonian gives
\begin{equation}
  H_k(\eta)
  =
  c_Z(\eta)Z+c_A(\eta)A+c_Z(\eta)\mathbb{I},
  \label{eq:H_ZA_with_const}
\end{equation}
where \(H_k\) denotes the Hamiltonian associated with the full momentum pair,
not a single oscillator mode. The coefficients are
\begin{equation}
  c_Z(\eta)
  =
  \frac{\omega_k^2(\eta)+\omega_0^2}{2\omega_0},
  \qquad
  c_A(\eta)
  =
  \frac{\omega_k^2(\eta)-\omega_0^2}{2\omega_0}.
  \label{eq:cZcA_general}
\end{equation}
The identity term contributes only a global phase and is omitted in the circuit
implementation. The term proportional to \(Z\) is number conserving, whereas
the term proportional to \(A\) creates and annihilates correlated
\((+\mathbf{k},-\mathbf{k})\) pairs.

Using Eq.~\eqref{eq:piecewise_frequency}, this gives
\begin{align}
  c_Z^{(\mathrm{dS})}(\eta)
  &=
  k-\frac{1}{k\eta^2},
  &
  c_A^{(\mathrm{dS})}(\eta)
  &=
  -\frac{1}{k\eta^2},
  \qquad \eta<\eta_e,
  \label{eq:cZcA-dS}
  \\
  c_Z^{(\mathrm{rad})}(\eta)
  &=
  k,
  &
  c_A^{(\mathrm{rad})}(\eta)
  &=
  0,
  \qquad \eta>\eta_e .
  \label{eq:cZcA-rad}
\end{align}
Thus the pair-creation term is active only in the de Sitter branch of the
sudden-transition model, while the radiation-era Hamiltonian is purely
number conserving in this fixed basis.

\subsection*{Discretization in conformal time}

For the numerical implementation it is convenient to use the dimensionless
conformal-time variable \(y\equiv k\eta.\)
For a given value of \(x=|k\eta_e|\), the transition occurs at \(y_e=k\eta_e=-x .\)
The evolution starts at \( y_i=-80,\) deep in the de Sitter branch, and ends at \(y_f=y_e+2,\) 
i.e. two dimensionless conformal-time units after the transition. The interval
\([y_i,y_f]\) is divided into \(N\) uniform steps,
\[
  y_n=y_i+n\Delta y,
  \qquad
  \Delta y=\frac{y_f-y_i}{N},
  \qquad n=0,\ldots,N .
\]
The Hamiltonian coefficients are evaluated at the midpoint of each interval,
\[
  y_{n+1/2}=\frac{y_n+y_{n+1}}{2}.
\]
The midpoint prescription also determines how the sudden transition is
treated numerically: if an interval straddles \(y_e\), the corresponding
Hamiltonian coefficients are selected according to whether
\(y_{n+1/2}<y_e\) or \(y_{n+1/2}\geq y_e\). The ambiguity associated with the
single interval containing the discontinuity decreases with increasing \(N\).

In terms of \(y=k\eta\), the evolution operator can be written in dimensionless
form. Since \(d\eta=dy/k\), the coefficients entering the \(y\)-evolution are
\(c_Z/k\) and \(c_A/k\). Using Eqs.~\eqref{eq:cZcA-dS}
and~\eqref{eq:cZcA-rad}, this gives
\[
  \tilde c_Z(y)=1-\frac{1}{y^2},
  \qquad
  \tilde c_A(y)=-\frac{1}{y^2},
  \qquad y<y_e ,
\]
and
\[
  \tilde c_Z(y)=1,
  \qquad
  \tilde c_A(y)=0,
  \qquad y\geq y_e .
\]
Here \(\tilde c_Z\equiv c_Z/k\) and \(\tilde c_A\equiv c_A/k\).

For each time slice the dimensionless Hamiltonian is therefore
\[
  \tilde H_k(y_{n+1/2})
  =
  \tilde c_Z(y_{n+1/2})Z
  +
  \tilde c_A(y_{n+1/2})A .
\]
The short-time evolution over the interval \([y_n,y_{n+1}]\) is approximated
using a second-order Strang splitting,
\begin{equation}
  U_k^{(n)}
  \simeq
  e^{-i\tilde c_Z(y_{n+1/2})Z\Delta y/2}
  e^{-i\tilde c_A(y_{n+1/2})A\Delta y}
  e^{-i\tilde c_Z(y_{n+1/2})Z\Delta y/2}.
  \label{eq:strang_step}
\end{equation}
The full time-ordered evolution is then approximated by
\begin{equation}
  U_k(y_f,y_i)
  \simeq
  U_k^{(N-1)}\cdots U_k^{(1)}U_k^{(0)},
  \label{eq:trotter_ordered_product}
\end{equation}
where \(U_k^{(0)}\) acts first on the initial state.

There are therefore two approximations in the simulator construction. First,
the continuous time-ordered evolution is replaced by a finite set of
midpoint-evaluated time slices. Second, within each slice the exponential of
the sum \(\tilde c_Z Z+\tilde c_A A\) is approximated by the second-order
product formula in Eq.~\eqref{eq:strang_step}. Since \(Z\) and \(A\) do not commute, this intra-slice product formula is
not exact at finite \(\Delta y\); the resulting deterministic approximation
error is the Trotter error.

The number of time slices \(N\) controls the accuracy of the time
discretization and the product-formula approximation. Increasing \(N\)
reduces these deterministic approximation errors, but also increases the
number of repeated circuit blocks and therefore the circuit depth. In the
hardware implementation discussed below, this depth growth is the main reason
for restricting the real-device demonstration to a shallow representative
block.

Unlike a one-shot implementation based on precomputed Bogoliubov coefficients,
the time-resolved construction uses only the Hamiltonian coefficients
\(\tilde c_Z(y_{n+1/2})\) and \(\tilde c_A(y_{n+1/2})\) at each step. Each
interval \([y_n,y_{n+1}]\) defines one short-time unitary, and the full
evolution is obtained by repeating the same logical block with different
rotation angles. This makes the simulation directly tied to the conformal-time
dynamics. The cost is that the circuit depth grows with the number of time
slices.

\subsection*{Digitization: 4-qubit encoding and operators}

To implement the dynamics on a gate-based quantum simulator, the Hilbert space
for the mode pair $(+k,-k)$ is truncated to at most one excitation per mode. A convenient
encoding, closely aligned with existing work on digital cosmological
simulations, uses two qubits per mode:
\begin{equation}
\begin{aligned}
  |01\rangle_{(+k)} &\equiv |0_+\rangle, &
  |10\rangle_{(+k)} &\equiv |1_+\rangle, \\
  |01\rangle_{(-k)} &\equiv |0_-\rangle, &
  |10\rangle_{(-k)} &\equiv |1_-\rangle .
\end{aligned}
\label{eq:mode_encoding}
\end{equation}
The global four-qubit states
\begin{equation}
\begin{aligned}
  |0101\rangle &\equiv |0_+,0_-\rangle,\\
  |1001\rangle &\equiv |1_+,0_-\rangle,\\
  |0110\rangle &\equiv |0_+,1_-\rangle,\\
  |1010\rangle &\equiv |1_+,1_-\rangle .
\end{aligned}
\label{eq:four_qubit_physical_basis}
\end{equation}
form the physical basis, while the remaining computational basis states are
unphysical and are not populated by the ideal dynamics.

On the four-dimensional physical subspace, the generators \(Z\) and \(A\)
defined in Eq.~\eqref{eq:ZA_def} are represented by finite matrices. These
matrices are embedded into the full \(16\)-dimensional four-qubit Hilbert
space and decomposed into sums of Pauli strings. We denote the resulting
four-qubit operators by \(Z_{\rm q}\) and \(A_{\rm q}\). They are
time-independent and are used at every Trotter step. The explicit Pauli
decompositions and representative pre-transpilation logical circuit are discussed in Appendix~\ref{app:pauli_decomposition}.

The Hamiltonian for the \(n\)-th time slice is represented on the qubit
register as
\begin{equation}
  H_k^{(n)}
  =
  c_Z(\eta_n) Z_{\rm q}
  +
  c_A(\eta_n) A_{\rm q}
  +
  c_Z(\eta_n)\mathbb{I}.
  \label{eq:qubit_slice_hamiltonian}
\end{equation}
The last term is proportional to the identity. At each Trotter step it
therefore contributes only a global phase, and the accumulated global phase
drops out of probability measurements. We therefore omit this term in the
circuit implementation. Since the operators \(Z_{\rm q}\) and
\(A_{\rm q}\) are fixed, the same logical circuit structure is reused at
each time slice, with only the rotation angles changing through
\(c_Z(\eta_n)\), \(c_A(\eta_n)\), and \(\Delta\eta_n\).

\subsection*{Observables and hardware diagnostics}
\label{subsec:observables}
The primary physical quantity extracted from the final state is the late-time
particle number in the two opposite-momentum modes. In the encoded basis, the particle numbers are
extracted from the measured probabilities as
\[
  n_{+\mathbf{k}} = P(1001)+P(1010),
  \qquad
  n_{-\mathbf{k}} = P(0110)+P(1010).
\]
For pair creation from the vacuum under the noiseless encoded Hamiltonian,
the single-particle probabilities \(P(1001)\) and \(P(0110)\) vanish. This
is because \(Z_{\rm q}\) preserves the total excitation number, while
\(A_{\rm q}\) changes it by two. Starting from the encoded vacuum
\(|0101\rangle\), the noiseless dynamics therefore remains in the
even-excitation sector spanned by \(|0101\rangle\) and \(|1010\rangle\).
Consequently, the produced particle number in the ideal truncated dynamics is
given by the pair probability \(P(1010)\).

Because the physical encoding occupies only a four-dimensional subspace of
the full sixteen-dimensional computational Hilbert space, noisy hardware
can populate unphysical basis states. We quantify this by the leakage
\[
  L =
  1-\left[
    P(0101)+P(1001)+P(0110)+P(1010)
  \right].
\]
This diagnostic is zero for ideal evolution and is used below to assess
the reliability of the IBM hardware results.

\section{Results}
\label{sec:results}

We now compare the time-resolved digital simulation with the analytic
sudden-transition benchmark derived in Sec.~\ref{sec:ds_radiation_model}.
The validation proceeds through a hierarchy of implementations. First, the
evolution is evaluated directly as a matrix-Trotter product in the
four-dimensional physical subspace, providing a high-resolution reference for
the truncated model. Second, the same time-resolved evolution is compiled into
the four-qubit encoding and evaluated with a noiseless Qiskit statevector
simulator~\cite{qiskit2024}, testing the Pauli decomposition and circuit
implementation. Third, the compiled circuit is sampled with the Qiskit Aer
shot-based simulator~\cite{qiskit2024}, isolating finite-shot fluctuations
before hardware noise is included. Finally, a shallow \(N=1\) IBM hardware
run is used as a feasibility test of the representative circuit block.

In the simulations below, the matrix-Trotter calculation uses \(N=2500\)
steps and serves as the highest-resolution truncated-subspace reference. The
noiseless Qiskit statevector simulation uses \(N=1000\); reducing the step
number from \(N=2500\) to \(N=1000\) was checked to leave the particle-number
curve unchanged within the line width of the validation plot. The finite-shot
Qiskit Aer simulations use \(N=500\), chosen to keep the sampled circuits
shallower while retaining the same qualitative trend. These finite-shot runs
are used to illustrate sampling fluctuations rather than as the
highest-precision convergence benchmark. The IBM hardware demonstration is
restricted to \(N=1\), where the circuit represents only a shallow
representative block rather than the full large-\(N\) time-resolved evolution.

For the numerical plots it is convenient to introduce the dimensionless
parameter \(x\equiv |k\eta_e|\). In terms of \(x\), the analytic result in
Eq.~\eqref{eq:sudden_particle_number} scales as \(n_k=1/(4x^4)\). Thus
smaller \(x\) corresponds to longer-wavelength modes that are more strongly
affected by the transition, while larger \(x\) corresponds to
shorter-wavelength modes with suppressed particle creation.

Since the four-qubit encoding excludes states with more than one excitation
per mode, the comparison with the full Bogoliubov result is controlled only
when the occupation is small. For a two-mode squeezed state with mean
occupation \(n_k\), the probability of occupying two or more pairs scales as
\[
  P_{\geq 2}
  =
  \left(\frac{n_k}{1+n_k}\right)^2
  =
  \mathcal{O}(n_k^2).
\]
Thus, for \(n_k<0.01\), the omitted multi-pair probability is below
\(10^{-4}\). Using Eq.~\eqref{eq:sudden_particle_number}, this corresponds
to \(x\gtrsim 2.24\). We therefore regard \(x\gtrsim 2.2\) as the most
conservative regime for quantitative comparison within the single-excitation
encoding. Smaller values of \(x\) are still useful for illustrating the
trend and for hardware feasibility tests, but they are interpreted less
strictly because the single-excitation truncation is less controlled.

\subsection*{Simulator validation}
\label{subsec:simulator_validation}

\begin{figure}[htbp]
    \centering
    \includegraphics[width=0.78\textwidth]{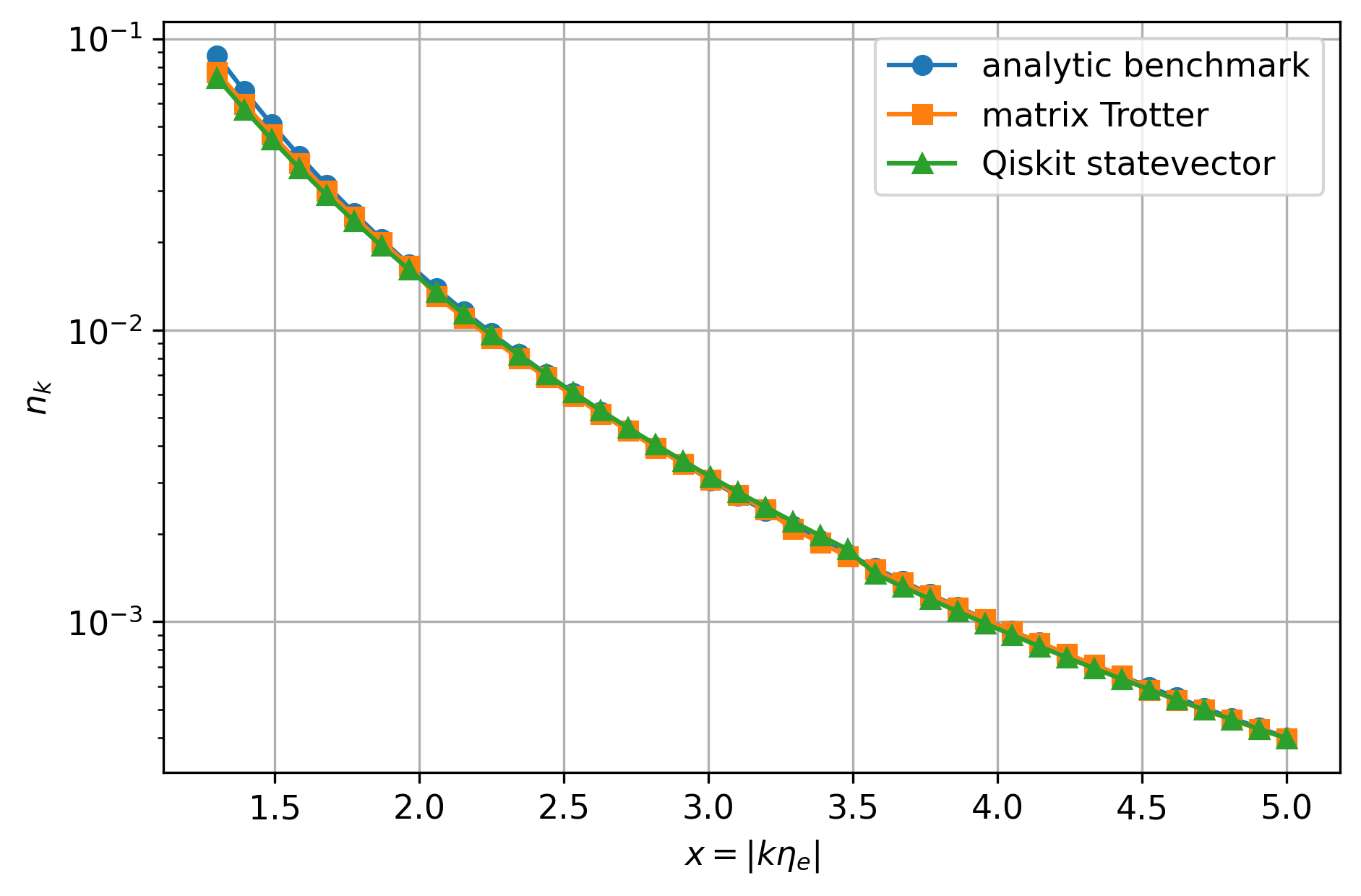}
    \caption{Time-resolved Trotterized quantum-circuit simulation. The analytic
benchmark is compared with the matrix-Trotter calculation and the Qiskit
statevector implementation.}
    \label{fig:qiskit_vs_analytic}
\end{figure}

We first validate the circuit construction in the noiseless setting.
Figure~\ref{fig:qiskit_vs_analytic} compares the analytic
sudden-transition benchmark, the matrix-Trotter result, and the Qiskit
statevector simulation. The statevector result reproduces the
matrix-Trotter calculation and follows the analytic curve in the valid
single-excitation regime. This confirms that the Pauli decomposition and
the four-qubit circuit implementation reproduce the intended time-sliced
Hamiltonian dynamics.

In the noiseless circuit simulation, the dynamics populates the vacuum state
\(|0101\rangle\) and the correlated pair state \(|1010\rangle\), while the
single-particle states \(|1001\rangle\) and \(|0110\rangle\) remain
numerically negligible. This confirms that the implemented dynamics produces
correlated \((+\mathbf{k},-\mathbf{k})\) pairs rather than spurious
single-mode excitations.

\subsection*{Finite-shot sampling}
\label{subsec:finite_shot_sampling}

We next evaluate the four-qubit time-resolved circuit with finite-shot
sampling using the Qiskit Aer simulator. The evolution remains noiseless, but
the exact statevector probabilities are replaced by sampled measurement
statistics. The resulting deviations therefore reflect finite-shot
fluctuations. The shot-based simulations use \(N=500\) conformal-time steps
and are performed with 8192, 32768, and 131072 shots.

\begin{figure}[htbp]
    \centering
    \includegraphics[width=0.78\textwidth]{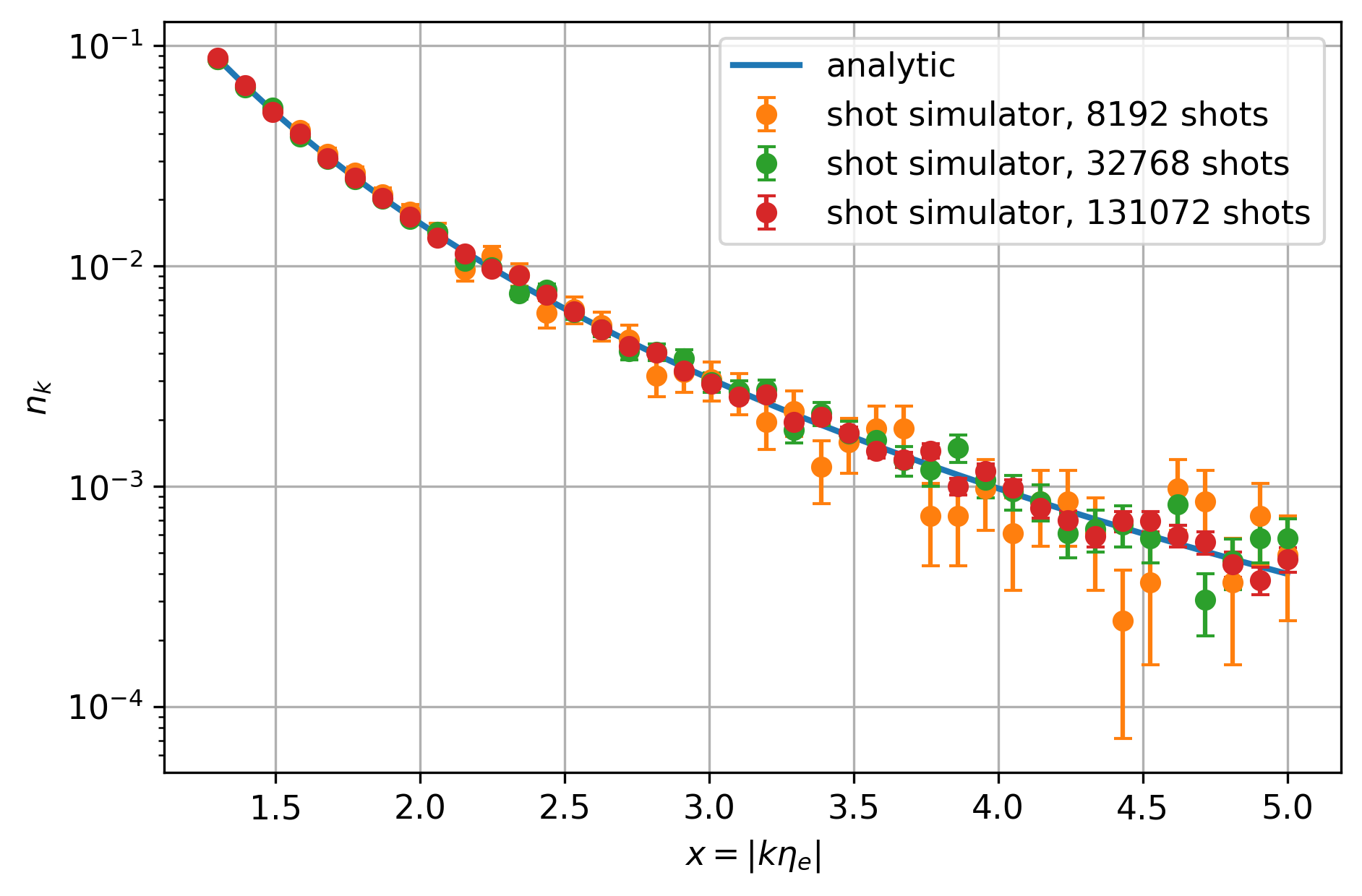}
    \caption{Finite-shot Qiskit Aer simulation of the time-resolved Trotterized
circuit for 8192, 32768, and 131072 shots. The sampled particle number
follows the analytic sudden-transition benchmark, with statistical
fluctuations decreasing as the number of shots increases. The largest
deviations occur at larger \(x\), where the particle-production probability
is smallest.}
    \label{fig:qiskit_shot_noise}
\end{figure}

Figure~\ref{fig:qiskit_shot_noise} compares the noiseless statevector
result with the finite-shot estimates. The sampled results follow the
statevector curve over the full range of \(x\), with larger fluctuations at
larger \(x\), where the particle number is smaller and fewer pair-creation
events are observed. Increasing the number of shots reduces the statistical
uncertainty, as expected.

The finite-shot simulations provide the statistical baseline for the
subsequent IBM hardware experiment. In the shot-based simulator, deviations
from the analytical result arise only from finite sampling. On real
hardware, additional effects enter, including gate errors, decoherence,
readout errors, and leakage out of the physical subspace. These effects are
analysed in the shallow hardware demonstration below.

\subsection*{Time-resolved production of particle pairs}
\label{subsec:time_resolved_trajectory}

To characterize the dynamical content of the time-resolved simulation, we
monitor the population of the encoded pair state during the evolution. The
initial state is the encoded in-vacuum \(|0101\rangle\), and within the
truncated physical subspace we define the fixed-basis pair occupation
\begin{equation}
    P_{\rm pair}(\eta) \equiv P_{1010}(\eta),
    \label{eq:pair_occupation_eta}
\end{equation}
where \(|1010\rangle\) denotes the state with one excitation in each of the
two modes \((+\mathbf{k},-\mathbf{k})\). During the de Sitter phase and the
transition, \(P_{\rm pair}(\eta)\) is a basis-dependent diagnostic of the
truncated dynamics rather than an invariant instantaneous particle number. In
the late radiation regime, where the reference basis coincides with the out
basis used in the analytic construction, its final value becomes the
corresponding out-particle estimate in the truncated Hilbert space.

\begin{figure}[htbp]
    \centering
    \includegraphics[width=0.82\textwidth]{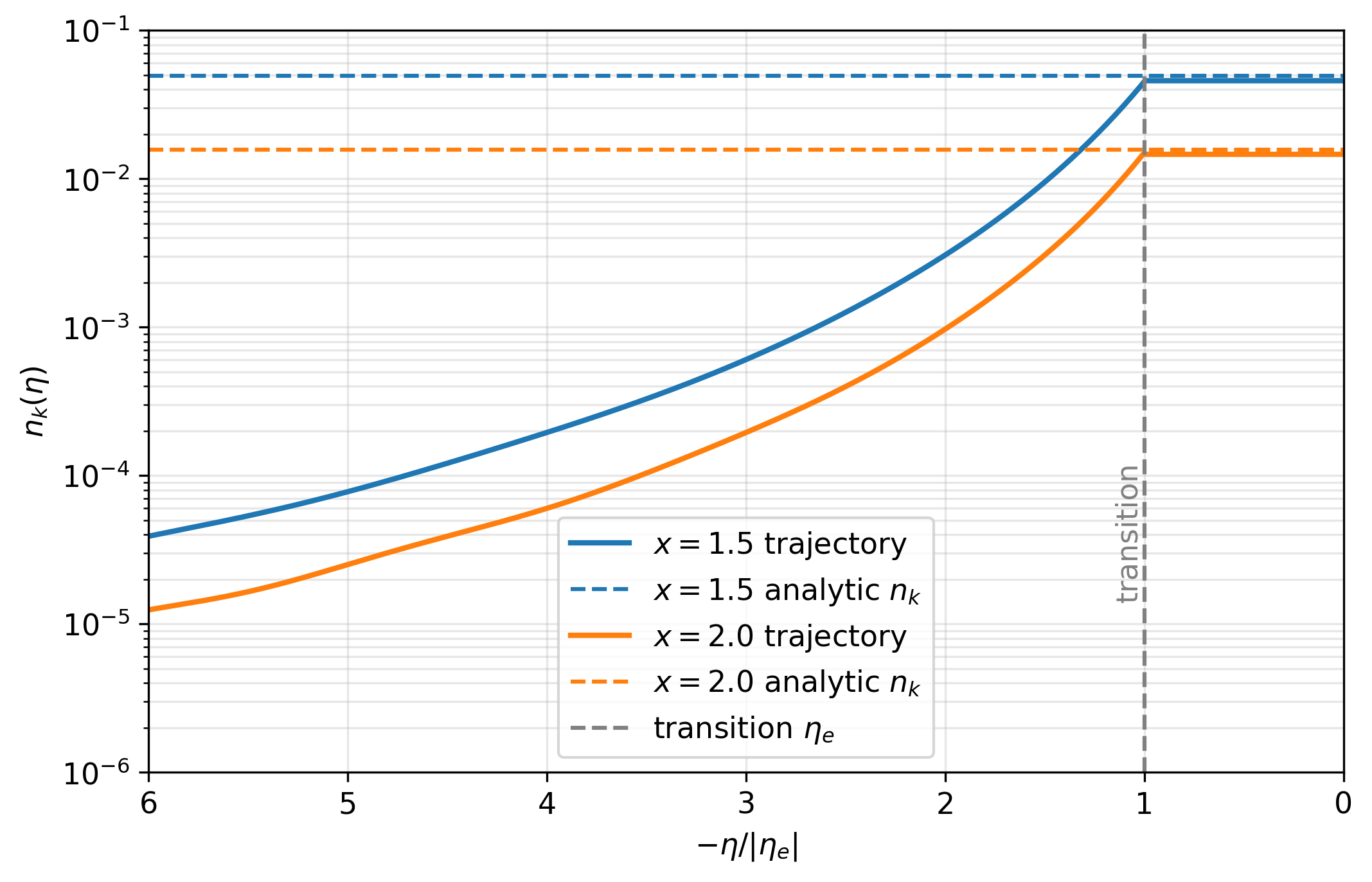}
    \caption{
    Time-resolved fixed-basis pair occupation
\(P_{\rm pair}(\eta)=P_{1010}(\eta)\) from the matrix-Trotter evolution for
two representative values of \(x=|k\eta_e|\). The dashed lines mark the
transition and the corresponding late-time analytic benchmarks.    }
    \label{fig:time_resolved_nk_eta}
\end{figure}

Figure~\ref{fig:time_resolved_nk_eta} shows \(P_{\rm pair}(\eta)\) for two
representative values of the dimensionless parameter $x=|k\eta_e|$. The horizontal axis is the
positive normalized conformal-time variable $-\eta/|\eta_e|$ and is inverted so
that the evolution proceeds from left to right. The transition from the
de~Sitter phase to the radiation phase occurs at $-\eta/|\eta_e|=1$.

The pair occupation remains strongly suppressed in the early de~Sitter region
and grows rapidly as the transition is approached. This behavior reflects the
non-adiabatic mixing generated by the time-dependent quadratic Hamiltonian. In
the radiation phase, the pair-creation coefficient vanishes in the fixed
radiation basis, and the occupation approaches an approximately constant
late-time value. The horizontal dashed lines in
Fig.~\ref{fig:time_resolved_nk_eta} show the analytic late-time benchmark
obtained from the sudden-matching Bogoliubov coefficient, \(n_k^{\rm an}=|\beta_k|^2={1}/{4x^4}.\) The larger final occupation for $x=1.5$ compared with $x=2.0$ is consistent
with the expected $x^{-4}$ scaling.

This diagnostic illustrates the distinction between the present time-resolved
construction and a one-shot implementation. A one-shot circuit based directly
on the final Bogoliubov coefficients can reproduce the asymptotic out-particle number, but it does not encode the intermediate history of the production
process. The time-resolved approach instead constructs the evolution from the
instantaneous Hamiltonian and applies the corresponding short-time propagators
sequentially. It therefore provides access to the build-up of the pair
occupation and identifies the transition region as the dominant source of the
produced particles.

\subsection*{IBM hardware demonstration}
\label{subsec:ibm_hardware}

As a final feasibility test, we executed a shallow representative block of the
time-resolved circuit on IBM quantum hardware. The full large-\(N\)
time-resolved simulation is too deep for current noisy devices, so the
hardware experiment was restricted to a single Trotter step, \(N=1\). This
\(N=1\) circuit should not be interpreted as a quantitative approximation to
the full time-resolved evolution. We therefore compute the noiseless
statevector output of the same logical \(N=1\) circuits used in the IBM runs;
this ideal \(N=1\) result provides the direct reference for the hardware data.
The analytic sudden-transition curve is retained only as the large-\(N\)
benchmark, while the hardware run is used to characterize executability,
readout error, gate noise, and leakage out of the physical subspace.

For the hardware demonstration we used the five-point parameter set
\[
    x = 1.3,\;1.5,\;1.8,\;2.0,\;2.2,
\]
with 4096 shots for each circuit. This range was chosen to keep the ideal \(N=1\) signal large enough to compare with the observed hardware noise floor.

 The shallow \(N=1\) circuits were executed on the \texttt{ibm\_fez}
backend, an IBM Quantum Heron r2 processor with 156 physical qubits. The
four logical qubits of the truncated Fock-space encoding were mapped to a
connected chain of four physical qubits. At the time of execution, the
backend-reported median two-qubit error rate was \(2.80\times 10^{-3}\),
and the median readout error was \(1.49\times 10^{-2}\).

These calibration values are relevant for interpreting the hardware data:
the readout error is already comparable to the particle-number signal in the
large-\(x\) regime, while accumulated gate errors are significant for
transpiled circuits with depth close to 100 and about 160 gates. Increasing
to \(N=2\) approximately doubled the transpiled depth and gate count; an
exploratory \(N=2\) run showed larger leakage and distorted particle-number
estimates. We therefore restrict the reported hardware study to \(N=1\).

To reduce readout bias, we applied matrix-free measurement mitigation
(M3)~\cite{nation2021m3}. M3 avoids constructing the full assignment
matrix and instead performs the correction in the subspace spanned by the
observed noisy bitstrings. In the present work it is used only as a
readout-error mitigation method; coherent gate errors, decoherence, and
Trotter-step errors are not corrected by M3.

To mitigate circuit-level noise, we also applied zero-noise extrapolation
(ZNE)~\cite{temme2017zne,li2017zne,kandala2019zne}. In ZNE, the observable is
estimated at several amplified noise levels and extrapolated to the
zero-noise limit. Here we used noise factors \((1,1.5,2)\) and a linear
extrapolator. Unlike M3, which acts only on readout errors, ZNE can partially
mitigate gate and decoherence errors. Its reliability, however, depends on
the quality of the noise-scaling procedure and on the assumption that the
measured observable varies smoothly over the extrapolation range. The narrow
noise range used here, \(1\times\)--\(2\times\), makes the linear
extrapolation particularly sensitive to statistical fluctuations,
contributing to the large ZNE uncertainties.

The numerical hardware results are summarized in
Table~\ref{tab:ibm_hardware_results}. The table separates three references:
the full analytic benchmark \(n_k^{\rm an}=1/(4x^4)\), the ideal noiseless
output of the same logical \(N=1\) circuits used on hardware, and the IBM
hardware estimates. The ideal \(N=1\) statevector values are approximately
\(2.5\times 10^{-3}\) over the full range of \(x\), substantially below the
full analytic benchmark. These values show why the ideal \(N=1\) statevector
result is the direct reference for the hardware data.

The raw and M3-mitigated particle numbers remain clustered around
\(10^{-2}\), well above the ideal \(N=1\) reference. This indicates that the
measured occupation is dominated by a residual hardware noise floor rather
than by the ideal single-step circuit response. M3 reduces the leakage from
approximately \(16\%\)--\(17\%\) to \(14\%\)--\(15\%\), but has little effect
on the extracted \(n_k\). ZNE gives smaller extrapolated leakage diagnostics
and moves closer to the ideal \(N=1\) reference for some values of \(x\), but
the particle-number estimates have large uncertainties and are not stable
enough for quantitative spectrum reconstruction. The M3 uncertainties quoted
in the table should be interpreted as approximate shot-noise indicators rather
than full propagated mitigation uncertainties.

\begin{table}[htbp]
\centering
\caption{
IBM hardware results for the shallow \(N=1\) circuit with 4096 shots.
The noiseless \(N=1\) statevector output is the direct reference for the
hardware data, while \(n_k^{\rm an}=1/(4x^4)\) is the full analytic benchmark.
Raw and M3 error bars are approximate finite-shot estimates; ZNE uncertainties
are returned by the Estimator/ZNE workflow.
}
\label{tab:ibm_hardware_results}
\resizebox{\textwidth}{!}{%
\begin{tabular}{c c c c c c c c c}
\hline
$x$ &
$n_k^{\rm an}$ &
$n_k^{N=1,\rm sv}$ &
$n_k^{\rm raw}$ &
$n_k^{\rm M3}$ &
$n_k^{\rm ZNE}$ &
leakage raw &
leakage M3 &
leakage ZNE \\
\hline
1.3 &
0.0875 &
0.0026 &
$0.0173 \pm 0.0020$ &
$0.0175 \pm 0.0020$ &
$0.0047 \pm 0.0148$ &
0.1604 &
0.1363 &
0.0747 \\

1.5 &
0.0494 &
0.0026 &
$0.0134 \pm 0.0018$ &
$0.0134 \pm 0.0018$ &
$0.0057 \pm 0.0155$ &
0.1665 &
0.1399 &
0.0753 \\

1.8 &
0.0238 &
0.0025 &
$0.0151 \pm 0.0019$ &
$0.0150 \pm 0.0019$ &
$0.0065 \pm 0.0268$ &
0.1729 &
0.1422 &
0.0981 \\

2.0 &
0.0156 &
0.0025 &
$0.0144 \pm 0.0019$ &
$0.0144 \pm 0.0019$ &
$0.0051 \pm 0.0133$ &
0.1677 &
0.1407 &
0.0521 \\

2.2 &
0.0107 &
0.0025 &
$0.0190 \pm 0.0021$ &
$0.0191 \pm 0.0021$ &
$0.0295 \pm 0.0226$ &
0.1738 &
0.1478 &
0.0859 \\
\hline
\end{tabular}%
}
\end{table}

\begin{figure}[htbp]
  \centering
  \includegraphics[width=0.78\textwidth]{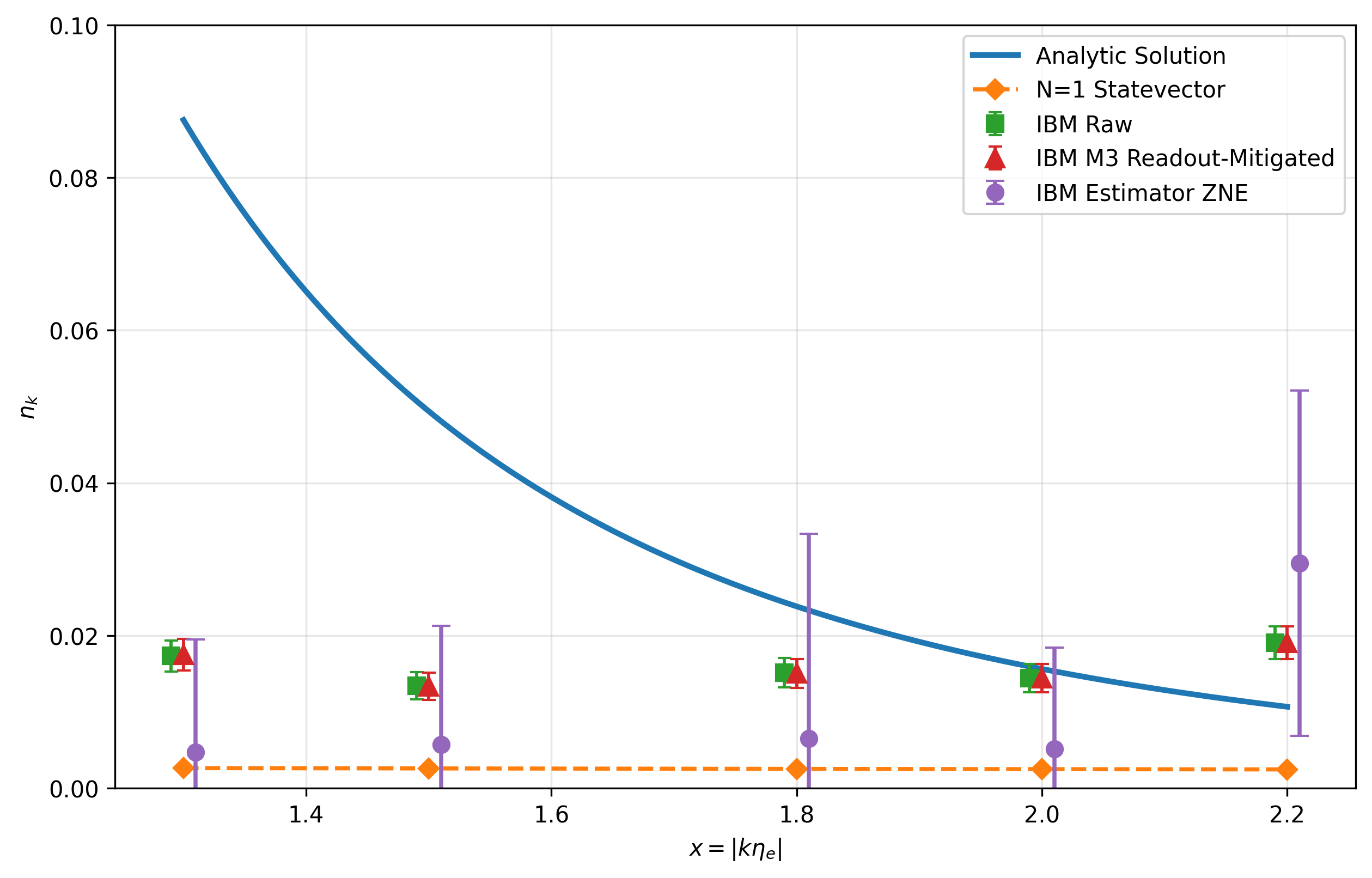}
  \caption{IBM hardware results for the shallow \(N=1\) circuit. The analytic
  curve is the large-\(N\) benchmark, while the noiseless \(N=1\) statevector
  points are the direct reference for the hardware data.}
  \label{fig:ibm_hardware}
\end{figure}

Figure~\ref{fig:ibm_hardware} visualizes the same comparison. The analytic
curve is retained as the large-\(N\) continuum benchmark, while the ideal
\(N=1\) statevector points show the noiseless target of the executed hardware
circuits. The raw and M3-mitigated data lie near a noise floor of order
\(10^{-2}\), and the ZNE estimates remain limited by large extrapolation
uncertainties.

These results show that, at the present circuit depth and noise level, the IBM
run is best interpreted as an execution test of the shallow \(N=1\) block and
not as a quantitative reconstruction of the particle spectrum.

\section{Discussion and Conclusion}
\label{sec:conclusion}

This work formulated cosmological particle creation in a sudden
de~Sitter--radiation transition as a time-resolved digital quantum simulation.
The essential difference from a one-shot Bogoliubov implementation is that the
circuit is generated from the time-dependent Hamiltonian coefficients at each
conformal-time slice, rather than from the final analytic Bogoliubov
coefficients. This makes the method less hardware-efficient, because the depth
grows with the number of time slices, but it exposes the dynamical build-up of
fixed-basis pair occupation during the non-adiabatic transition.

The simulator results show that this construction is internally consistent in
the controlled single-excitation regime. The matrix-Trotter calculation agrees
with the analytic sudden-transition benchmark within the expected finite-step
accuracy, and the noiseless Qiskit statevector simulation reproduces the
matrix-Trotter result. This separates two checks: the physical truncated
time-resolved evolution is tested against the analytic benchmark, while the
qubit encoding and Pauli decomposition are tested against the matrix evolution.
The Finite-shot Aer simulations then show the expected statistical convergence with
increasing shot number.

The IBM hardware run has a more limited role. Because current devices cannot
execute the large-\(N\) circuits required for the full time-resolved simulation,
the hardware experiment was restricted to a shallow \(N=1\) representative
block. The raw and M3-mitigated data show a noise floor of order \(10^{-2}\),
and the ZNE estimates are not stable enough to reconstruct the particle
spectrum quantitatively. The hardware results should therefore be interpreted
as a feasibility test of the compiled circuit block and as a diagnostic of
noise and leakage, not as a hardware verification of cosmological particle
creation.

The main limitations are the truncation to at most one excitation per mode, finite Trotter
resolution, circuit depth, two-qubit-gate errors, readout error, and leakage
into unphysical computational-basis states. The sudden transition was chosen
because it provides a closed-form benchmark; a natural next step is to replace
it by a smooth interpolation, where the same time-resolved construction can be
used without relying on closed-form Bogoliubov coefficients. Further extensions
include increasing the Fock cutoff and optimizing the circuit decomposition to
make larger-\(N\) simulations more accessible on future hardware.

Overall, the present work should be viewed as a controlled truncated-model
demonstration of Hamiltonian-based, time-resolved quantum simulation for
curved-spacetime particle creation. The full dynamics are accurately validated
on quantum-circuit simulators, while current IBM hardware remains limited to
shallow proof-of-principle circuit execution.

\appendix
\section{Pauli decomposition of the four-qubit generators}
\label{app:pauli_decomposition}

In this appendix we give the explicit four-qubit representation of the
time-independent operators used in the circuit construction. The physical
basis is ordered as
\[
  |0101\rangle,\quad |1001\rangle,\quad |0110\rangle,\quad |1010\rangle,
\]
corresponding respectively to
\[
  |0_+,0_-\rangle,\quad |1_+,0_-\rangle,\quad
  |0_+,1_-\rangle,\quad |1_+,1_-\rangle .
\]

On this four-dimensional physical subspace, the number-conserving generator
\(Z\) defined in Eq.~\eqref{eq:ZA_def} is represented by
\begin{equation}
  Z_{\rm phys}
  =
  \begin{pmatrix}
    0 & 0 & 0 & 0 \\
    0 & 1 & 0 & 0 \\
    0 & 0 & 1 & 0 \\
    0 & 0 & 0 & 2
  \end{pmatrix}.
  \label{eq:Z_phys_matrix}
\end{equation}
The entries \(0,1,1,2\) are the total occupation numbers of the two-mode
states. The pair-creation generator \(A\) is represented by
\begin{equation}
  A_{\rm phys}
  =
  \begin{pmatrix}
    0 & 0 & 0 & 1 \\
    0 & 0 & 0 & 0 \\
    0 & 0 & 0 & 0 \\
    1 & 0 & 0 & 0
  \end{pmatrix}.
  \label{eq:A_phys_matrix}
\end{equation}
Thus \(A_{\rm phys}\) couples the encoded vacuum state \(|0101\rangle\) to
the encoded pair state \(|1010\rangle\), while the single-particle states
\(|1001\rangle\) and \(|0110\rangle\) are left uncoupled.

To embed these matrices into the full four-qubit Hilbert space, we use the
two-qubit encoding of each mode. For the \(+\mathbf{k}\) mode,
\(|01\rangle_{01}\) represents the vacuum and \(|10\rangle_{01}\) represents
one excitation. Therefore the occupation projector for the \(+\mathbf{k}\)
mode is
\begin{equation}
  N_+
  =
  |10\rangle_{01}\langle 10|
  =
  \frac{1}{4}
  \left(\mathbb{I}-\sigma_0^z\right)
  \left(\mathbb{I}+\sigma_1^z\right).
  \label{eq:Nplus_projector}
\end{equation}
Similarly, for the \(-\mathbf{k}\) mode,
\begin{equation}
  N_-
  =
  |10\rangle_{23}\langle 10|
  =
  \frac{1}{4}
  \left(\mathbb{I}-\sigma_2^z\right)
  \left(\mathbb{I}+\sigma_3^z\right).
  \label{eq:Nminus_projector}
\end{equation}
The embedded four-qubit number operator is therefore
\begin{equation}
  Z_{\rm q}=N_+ + N_- .
  \label{eq:Zq_projector_form}
\end{equation}
Equivalently,
\begin{align}
  Z_{\rm q}
  =
  \frac{1}{2}\mathbb{I}
  &-\frac{1}{4}\sigma_0^z
  +\frac{1}{4}\sigma_1^z
  -\frac{1}{4}\sigma_2^z
  +\frac{1}{4}\sigma_3^z
  \nonumber\\
  &-\frac{1}{4}\sigma_0^z\sigma_1^z
  -\frac{1}{4}\sigma_2^z\sigma_3^z .
  \label{eq:Zq_pauli}
\end{align}
Here and below, tensor products with identity operators on the remaining
qubits are implicit.

The pair-creation part must map the encoded vacuum state \(|0101\rangle\) to
the encoded pair state \(|1010\rangle\), and the pair-annihilation part must
perform the inverse map. Define the transition operators
\begin{equation}
  C_+
  =
  |10\rangle_{01}\langle 01|,
  \qquad
  C_-
  =
  |10\rangle_{23}\langle 01| .
  \label{eq:Cplus_Cminus_def}
\end{equation}
Then the embedded pair generator is
\begin{equation}
  A_{\rm q}
  =
  C_+ C_-
  +
  C_+^\dagger C_-^\dagger .
  \label{eq:Aq_transition_operator}
\end{equation}
Equivalently,
\begin{equation}
  A_{\rm q}
  =
  |1010\rangle\langle 0101|
  +
  |0101\rangle\langle 1010| .
  \label{eq:Aq_projector_form}
\end{equation}
This is the four-qubit embedding of \(A_{\rm phys}\).

Using
\begin{equation}
  |1\rangle\langle 0|
  =
  \frac{1}{2}\left(\sigma^x-i\sigma^y\right),
  \qquad
  |0\rangle\langle 1|
  =
  \frac{1}{2}\left(\sigma^x+i\sigma^y\right),
  \label{eq:raising_lowering_pauli}
\end{equation}
one obtains the Pauli-string decomposition
\begin{align}
  A_{\rm q}
  =
  \frac{1}{8}\Big(
  &\sigma_0^x\sigma_1^x\sigma_2^x\sigma_3^x
  +\sigma_0^x\sigma_1^x\sigma_2^y\sigma_3^y
  -\sigma_0^x\sigma_1^y\sigma_2^x\sigma_3^y
  +\sigma_0^x\sigma_1^y\sigma_2^y\sigma_3^x
  \nonumber\\
  &+\sigma_0^y\sigma_1^x\sigma_2^x\sigma_3^y
  -\sigma_0^y\sigma_1^x\sigma_2^y\sigma_3^x
  +\sigma_0^y\sigma_1^y\sigma_2^x\sigma_3^x
  +\sigma_0^y\sigma_1^y\sigma_2^y\sigma_3^y
  \Big).
  \label{eq:Aq_pauli}
\end{align}
Restricted to the physical subspace, \(Z_{\rm q}\) and \(A_{\rm q}\)
reproduce the matrices \(Z_{\rm phys}\) and \(A_{\rm phys}\) in
Eqs.~\eqref{eq:Z_phys_matrix} and~\eqref{eq:A_phys_matrix}. In particular,
\[
  \langle 0101|A_{\rm q}|1010\rangle
  =
  \langle 1010|A_{\rm q}|0101\rangle
  =
  1,
\]
while the matrix elements connecting the single-particle states to the
vacuum or pair state vanish.

\begin{figure}[htbp]
  \centering
  \includegraphics[width=\textwidth]{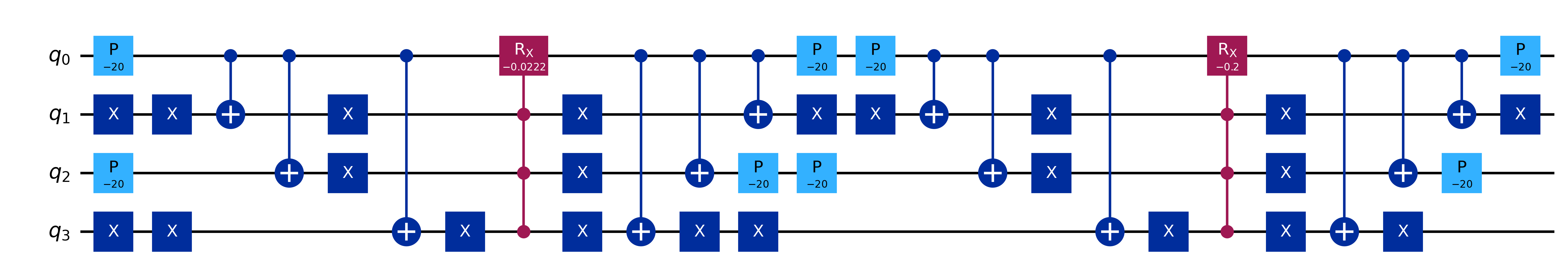}
  \caption{Representative pre-transpilation logical circuit for two Trotter
  steps of the four-qubit implementation.}
  \label{fig:logical_circuit_appendix}
\end{figure}

Figure~\ref{fig:logical_circuit_appendix} shows a representative logical
circuit for two Trotter steps of the time-resolved simulation, generated from
the Pauli-string implementation of \(Z_{\rm q}\) and \(A_{\rm q}\). The
initial \(X\) gates on \(q_1\) and \(q_3\) prepare the encoded fixed-basis
vacuum \(|0101\rangle\) from the hardware default state \(|0000\rangle\).
Each Trotter block implements the second-order Strang splitting in
Eq.~\eqref{eq:strang_step}: a half-step under \(Z_{\rm q}\), a full step under
\(A_{\rm q}\), and a second half-step under \(Z_{\rm q}\). The \(A_{\rm q}\)
part is implemented by the central \(R_X\) rotations together with the
surrounding CNOT and \(X\) gates. The rotation angles vary from one block to
the next because they are determined by the time-dependent coefficients
\(\tilde c_A(y_{n+1/2})\). The figure is shown before hardware transpilation;
the IBM circuits used in the hardware runs are obtained by mapping this
logical circuit to the backend-native gate set and qubit connectivity. After transpilation, even the \(N=1\) hardware circuits have depths close to
100, illustrating the hardware cost of the logical block.

The embedding outside the physical subspace is not unique. The choice above
is the natural embedding induced by the two-qubit mode encoding and the
transition operators in Eq.~\eqref{eq:Cplus_Cminus_def}. The ideal dynamics
generated by \(Z_{\rm q}\) and \(A_{\rm q}\) preserves the physical subspace;
population outside this subspace in hardware runs is therefore interpreted
as leakage caused by noise.

\bibliography{PCQC_References}
\end{document}